\documentclass[reprint,amsmath,amssymb,aps,prb]{revtex4-1}
\usepackage{graphicx}
\usepackage{dcolumn}
\usepackage{bm}
\usepackage{natbib}
\usepackage{color}

\begin{document}

\preprint{APS/123-QED}

\title{Brillouin light scattering spectroscopy of parametrically excited dipole-exchange magnons}

\author{A.~A.~Serga}
\email{serga@physik.uni-kl.de}
\affiliation{%
Fachbereich Physik and Forschungszentrum OPTIMAS, Technische Universit\"at Kaiserslautern, 67663 Kaiserslautern, Germany}%
\author{C.~W.~Sandweg}
\affiliation{%
Fachbereich Physik and Forschungszentrum OPTIMAS, Technische Universit\"at Kaiserslautern, 67663 Kaiserslautern, Germany}%
\author{V.~I.~Vasyuchka}
\affiliation{%
Fachbereich Physik and Forschungszentrum OPTIMAS, Technische Universit\"at Kaiserslautern, 67663 Kaiserslautern, Germany}%
\author{M.~B.~Jungfleisch}
\affiliation{%
Fachbereich Physik and Forschungszentrum OPTIMAS, Technische Universit\"at Kaiserslautern, 67663 Kaiserslautern, Germany}%
\author{B.~Hillebrands}
\affiliation{%
Fachbereich Physik and Forschungszentrum OPTIMAS, Technische Universit\"at Kaiserslautern, 67663 Kaiserslautern, Germany}%
\author{A.~Kreisel}
\affiliation{%
Institut f\"ur Theoretische Physik, Universit\"at Frankfurt, 60438 Frankfurt, Germany}%
\author{P.~Kopietz}
\affiliation{%
Institut f\"ur Theoretische Physik, Universit\"at Frankfurt, 60438 Frankfurt, Germany}%
\author{M.~P.~Kostylev}
\affiliation{%
School of Physics, M013, University of Western Australia, Crawley, WA 6009, Australia
}%

\date{\today}

\begin{abstract}
The spectral distribution of parametrically excited dipole-exchange magnons in an in-plane magnetized epitaxial film of yttrium-iron garnet was studied by means of frequency- and wavevector-resolved Brillouin light scattering spectroscopy. The experiment was performed in a parallel pumping geometry where an exciting microwave magnetic field was parallel to the magnetizing field. It was found that for both dipolar and exchange spectral areas parallel pumping excites the lowest volume magnon modes propagating in the film plane perpendicularly to the magnetization direction. In order to interpret the experimental observations, we used a microscopic Heisenberg model that includes exchange as well as dipole-dipole interactions to calculate the magnon spectrum and construct the eigenstates. As proven in our calculations, the observed magnons are characterized by having the highest possible ellipticity of precession which suggests the lowest threshold of parametric generation. Applying different pumping powers we observe modifications in the magnon spectrum that are described theoretically by a softening of the spin stiffness.
\end{abstract}

\maketitle

\section{Introduction}
Parallel pumping, the parametric excitation of spin waves by means of a linearly polarized microwave magnetic field parallel to an external bias magnetic field, was first described in 1960.\cite{Morgenthaler60, Schloemann60} The technique has been used for the last five decades to investigate a number of interesting phenomena such as the excitation of magnetoelastic waves in ferromagnets,\cite{Schloemann69} spin-wave parametric instability,\cite{Schloemann60, Wiese94} high density magnon gases and condensates,\cite{Demokritov06, Demidov08, Kloss10, Neumann08} parametric amplification of spin-wave solitons and bullets,\cite{Kostylev_soliton, Patton_soliton, Serha_bullet} spin pumping and the inverse spin Hall effect.\cite{Sandweg11, KurebayashiNatureMat11, KurebayashiAPL11, Ando11} While the main principles of parametric pumping have been revealed in experiments with bulk ferrimagnetic samples the more recent research is focused on the application of this technique to ferro- and ferrimagnetic films. Thus, the understanding of the peculiarities of parametric excitation in the application to magnetic films is of paramount importance for the correct interpretation of many interesting effects (see, for example, Refs.~\onlinecite{Sandweg11, KurebayashiAPL11, Ciubotaru11} focused on microwave driven ferrite-platinum bi-layers and nano-contacts).

Parallel pumping has two essential advantages in comparison with the direct excitation of spin waves. Firstly, when spin waves are directly driven by a microstrip antenna using an alternating Oersted field, the wavenumber of the effectively excited spin waves is rather small, in the range of $10^{3}$~rad/cm due to the finite width of the antenna. Therefore, direct excitation of large-wavenumber spin waves can hardly be realized. Secondly, parallel pumping is a threshold process, i.e. the driving field amplitude $h$ should exceed a certain threshold value $h_\mathrm{crit}$, which is dependent on the magnetic losses in the material, in order to start the parametric generation. Once this threshold has been overcome, the amplitude of the parametric spin waves grows exponentially in time until higher order nonlinear processes limit this amplitude to a certain level.\cite{Lvov, Gurevich_Melkov} With this method it is possible to achieve much higher spin-wave intensities in comparison to the direct linear excitation of spin waves with a microstrip antenna.

\begin{figure}[tb]
\begin{center}
\scalebox{1}{\includegraphics[width=8.5 cm]{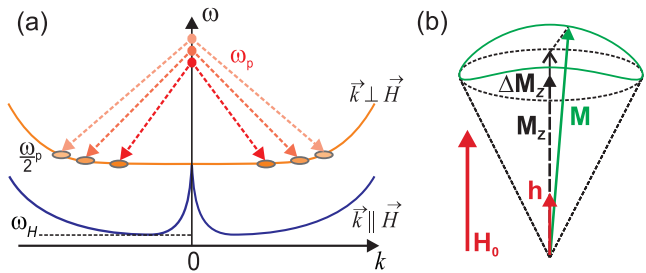}}
\end{center}
\caption{\label{Fig1} (Color online) (a) Spectral position of the parametrically amplified magnon modes using the pumping frequency $\omega_\mathrm p$. (b) The elliptical precession of the magnetization $\bf M$ allows the excitation of magnon modes by an alternating magnetic field $\bf h$ parallel to the external bias magnetic field ${\bf H}_0$.}
\end{figure}

During the process of parallel parametric amplification, spin waves are excited at half of the initial microwave pumping frequency $\omega_\mathrm{p}$. In terms of energy quanta, a microwave photon splits into two magnons which are the quasiparticles of the dynamic magnetization. From the laws of the conservation of energy and momentum and from the fact that for the case of weakly localized pumping the photon wavenumber is negligibly small,~\cite{Melkov_JETP} it then follows that the pumping process creates pairs of magnons with the same frequency ${\omega_{p}}/{2}$ and with opposite wavevectors $\bf k$ and $-\bf k$, as shown in Fig.~\ref{Fig1}~(a). Different magnon groups with the same frequency ${\omega_{p}}/ {2}$ are parametrically driven at the same time, and only the one with the lowest damping survives.\cite{Lvov, restoration1, restoration2} The determination of the spectral position of this dominant group is of crucial importance not only for the investigation of the primarily pumped magnons but for all the magnon groups which acquire their energy in subsequent scattering processes from the dominant group.

Here we present a direct observation of the dominant group for different pumping frequencies and powers in a wide wavevector range up to $1.6\times 10^{5}$~rad/cm using Brillouin light scattering (BLS) spectroscopy. The experiments are performed on a single-crystal epitaxial film of yttrium-iron garnet (YIG).\cite{SagaYIG, YIG_Magnonics} We show experimentally that the primarily excited spin waves of the dominant group are always located at the 90$^\circ$ branch of the backward volume magnetostatic wave (BVMSW) having the lowest energy. Previous works on this topic were carried out by Kabos~\textit{et~al.},\cite{Kabos97, Wettling83} but only for dipolar-dominated spin waves with a wavenumber up to $4\times 10^{4}$~rad/cm. In our present work we complete the picture by obtaining more details for the dipolar-dominated waves as well as by exploring the parametric processes involving the exchange-dominated BVMSW. By changing the pumping frequency $\omega_\mathrm{p}$ in sufficiently small step sizes, as indicated in Fig.~\ref{Fig1}~(a), it is possible to move along the dispersion branch of the BVMSW and to determine the spectral position of the dominant group. In addition, calculations of the magnon dispersion based on an effective spin model for YIG have been performed which are in excellent agreement with the experimentally obtained results.

This paper is organized as follows: Sec.~\ref{sec:ppa} provides the reader with basic information about the technique of parallel parametric amplification. Calculations of the relevant magnon modes are performed in Sec.~\ref{sec:theory}. In Sec.~\ref{sec:setup} details of the combined BLS/microwave experiment are given, especially the wide wavevector selectivity. Sec.~\ref{sec:expdat} shows the experimental results which clarify the position of the dominant group for different pumping frequencies and powers. These experimental data are supported by theoretical investigations of the spin-wave dispersion. Finally, in Sec.~\ref{sec:conclusion}, we summarize our results and present our conclusions.

\section{Parallel parametric amplification}
\label{sec:ppa}

A key condition for the parametric amplification is ellipticity of magnetization precession. The circular precession implies that the tip of the magnetization vector goes around a circle lying in the plane perpendicular to the applied field. In this case the component of the magnetization along the external bias magnetic field is constant:
\begin{equation}
 M_\mathrm{z}=\mathrm{const}.
\end{equation}
The parametric amplification can be considered as a response of the spin system to the application of a weak alternating magnetic field parallel to the bias magnetic field ${\bf H}_\mathrm{0}$, which is oscillating with a precession frequency $2\omega_\mathrm{0}$ which is twice the eigenfrequency of the spin wave $\omega_\mathrm{0}$ we intend to excite parametrically:
\begin{equation}
 h(t)=h_\mathrm{0}\cos(2\omega_\mathrm{0}t).
\end{equation}
Therefore, for the purely circular precession, the average energy pumped into the system vanishes:
\begin{equation}
\left\langle \Delta E\right\rangle \propto \left\langle \textbf{h}\cdot\textbf{M} \right\rangle =
\left\langle h \right\rangle  M_\mathrm{z} = h_\mathrm{0}M_\mathrm{z}
\int^{T_\mathrm{0}}_{0}\frac{dt}{T_0}\cos(2\omega_\mathrm{0}t)\equiv 0,
\end{equation}
where $\bf{M}$ denotes the magnetization and $T_\mathrm{0}$ denotes the time for one cycle of precession.

An elliptical precession implies that this circle is squeezed in some direction in the plane which is perpendicular to the $z$-axis.
As shown in Fig.~\ref{Fig1}~(b), in this case the $M_\mathrm{z}$ component of magnetization oscillates with double frequency:
\begin{equation}
 M_\mathrm{z}=\mathrm{const}.+m_\mathrm{z,0} \cos(2\omega_\mathrm{0}t).
\end{equation}
This allows absorption of microwave energy by the spin system:
\begin{equation}
\left\langle \Delta E\right\rangle \propto
\int^{T_\mathrm{0}}_{0}\frac{dt}{T_0}\cos^{2}(2\omega_\mathrm{0}t)\neq 0.
\end{equation}

For a bulk magnetic material the dynamic demagnetizing field along the direction of propagation of the plane spin wave gives rise to the ellipticity of the magnetization precession. In this case, the threshold for the parallel pumping is given by a simple analytical formula:\cite{Gurevich_Melkov}
\begin{equation}
h_\mathrm{crit}=\min \left\{ \frac{\omega_\mathrm{p} \Delta
H_{\bf k}}{\omega_\mathrm{M}\sin^{2}\vartheta_{\bf k}} \right\},
\end{equation}
where $\Delta H_{\bf k} = 2 \Gamma_{\bf k}/ \gamma$, $\omega_\mathrm{M}=\gamma 4 \pi M_\mathrm{s}$, and $\vartheta_{\bf k}$ is the angle between the propagation direction of the parametrically amplified spin waves with wavevector $\bf k$ and the external bias magnetic field. $\Gamma_{\bf k}$ denotes the relaxation frequency of this magnon mode, $\gamma$ is the gyromagnetic ratio, and $M_\mathrm{s}$ is the saturation magnetization.
Importantly, from this expression it follows that the threshold field $h_\mathrm{crit}$ is inversely proportional to the ellipticity $\epsilon_{\bf k}$, which is defined by
\begin{equation}
\epsilon_{\bf k} = 1 - \frac{m_{x_{\bf k}}^2}{m_{y_{\bf k}}^2} ,
\label{eq:ellipticity}
\end{equation}
where $m_x$ and $m_y$ are the amplitudes of the dynamic components of the total magnetization vector ${\bf M}=({\bf m}_x, {\bf m}_y, {\bf M}_z)$, and ${\bf y} \bot {\bf k}$.
Consequently, $h_\mathrm{crit}$ reaches the smallest possible value at $\vartheta_{\bf k}={\pi}/{2}$ when the ellipticity reaches its maximum.\cite{Gurevich_Melkov, Wiese94}

In the confined geometry of thin films the picture of the dynamic demagnetizing field is more complicated. The out-of-plane component of the dynamic demagnetizing field, caused by the presence of the two film surfaces, ``compresses'' the precession cone in the direction of the film plane. As a result, the value of $h_\mathrm{crit}$ becomes dependent not only on the direction (as for the bulk materials), but also on the value of the in-plane wavenumber. In addition, the finite film thickness results in quantization of the total magnon wavenumber in the direction perpendicular to the film surface. A simple analytical expression can easily be obtained for the mode of the uniform precession (the wave with the vanishing value of the  total wavenumber) $h_\mathrm{crit} \propto 1 / \eta_{{\bf k},n}$, where
\begin{equation}
\eta_{{\bf k},n} =  \left(1 - \frac{m_{x_{\bf k},n}^2}{m_{y_{\bf k},n}^2}\right)\left(\frac{m_{y_{\bf k},n}}{m_{x_{\bf k},n}}\right) .
\end{equation}
In this work we extend the validity of this formula for the entire range of the wavevectors parametrically excited in the films. Under this assumption, and with the calculation shown in the next section, we find that in the case of intermediate film thicknesses (when $m_{x_{\bf k},n} \lesssim m_{y_{\bf k},n}$) no qualitative difference between $\eta$ and $\epsilon$ exists. Therefore, in the following, we restrict ourself to the classical definition of ellipticity given by Eq.~\ref{eq:ellipticity}. However, in contrast to bulk materials\cite{Gurevich_Melkov}, in this work the discrete character of the magnon spectrum is also taken into account.

Another particular type of parametric amplification is perpendicular pumping. It is characterized by the application of an alternating pumping field in the plane perpendicular to the direction of the equilibrium magnetization, i.e, the alternating magnetization induced by the pumping field is always perpendicular to the external bias magnetic field.\cite{Suhl57} In this case, the amplification is caused by the parametric decay of the uniform precession mode, driven out-of-resonance (i.e. at $2 \omega_0$). In our work we use the microwave Oersted field of a microstrip transducer for the parametric pumping. The Oersted field of such a transducer has two components: one in-plane and one out-of-plane. The in-plane component is parallel to the bias field and thus is responsible for the parallel pumping process. Similarly, the out-of-plane component of the alternating Oersted field is responsible for a possible perpendicular pumping process\cite{NeumannAPL09} which we want to avoid in our work. Fortunately, the in-plane component of the Oersted field is maximized below the longitudinal axis of the microstrip line and, the out-of-plane component is mostly localized near the microstrip edges. Therefore, by using a sufficiently small laser spot size and probing the area near the longitudinal axis one can exclude the contribution of the perpendicular pumping mechanism from the parametric process.

\section{Theoretical investigations of the magnon spectrum}
\label{sec:theory} In order to find the spectral positions of the parametrically injected magnons in a film sample, we used a recently developed microscopic approach~\cite{KreiselEPJ} for calculating the magnon energies as well as the ellipticity of the precession. Our starting point is a microscopic Hamiltonian that describes the properties of the relevant magnon modes in YIG. The model for our magnetic films of a finite thickness $d$ contains both exchange and dipole-dipole interactions and is completed by a Zeeman term that takes into account the external bias magnetic field ${\bf H}_0$:

\begin{align}
{\cal{H}}&=-\frac 12 \sum_{ij} J_{ij} {\bf S}_i \cdot {\bf S}_j -\mu {\bf H}_\mathrm{0}\cdot \sum_i{\bf S}_i\notag\\
&\hspace{-.5cm}\phantom{=}-\frac 12 \sum_{ij,i\neq j}\frac{\mu^2}{|{\bf R}_{ij}|^3} \left[3 ({\bf S}_i\cdot\hat{\bf R }_{ij})({\bf S}_j\cdot\hat{\bf R }_{ij}) -{\bf S}_i \cdot {\bf S}_j\right].
 \label{eq:hamiltonian}
\end{align}
The spin operators ${\bf S}_i$ are normalized such that ${\bf S}_i^2=S(S+1)$ with an effective total spin $S$ per lattice site. The sums run over the sites  ${\bf R}_i$ of a cubic lattice with spacing $a=12.376\,\text{\AA}$, and $\hat{\bf R}_{ij}={\bf R}_{ij}/|{\bf R}_{ij}|$ are unit vectors in the direction of ${\bf R}_{ij}={\bf R}_i-{\bf R}_j$. The relevant parameters are the exchange interactions $J_{ ij}=J$ of neighboring spins and the magnetic moment $\mu=g\mu_B$, where $g$ is the effective $g$-factor and $\mu_B$ is the Bohr magneton. In order to proceed using a compact notation we introduce the dipolar tensor $D_{ij}^{\alpha \beta}=D^{\alpha\beta}({\bf R}_i-{\bf R}_j)$,
\begin{align}
 D_{ij}^{\alpha\beta}&=(1-\delta_{ij})\frac{\mu^2}{|{\bf R}_{ij}|^3}\left[3\hat{R}_{ij}^\alpha\hat{R}_{ij}^\beta-\delta^{\alpha\beta}\right]
\label{eq:defdip}
\end{align}
and rewrite our effective Hamiltonian (\ref{eq:hamiltonian}) as
\begin{equation}
 {\cal{H}}=-\frac 12 \sum_{ij}\sum_{\alpha\beta} \left[J_{ij}\delta^{\alpha\beta}+D_{ij}^{\alpha\beta}\right] S_i^\alpha S_j^\beta -h\sum_i S_i^z ,
\end{equation}
with the z-axis of the frame of reference pointing along the direction of the external magnetic field ${\bf H}_0=h/\mu {\bf e}_z$. We assume that the continuous film is magnetically saturated by the magnetic field ${\bf H}_0$ in its plane and the ferromagnetic magnetic order is perfect. This allows us to expand the Hamiltonian in terms of bosonic operators describing fluctuations around the classical ground state. Inserting the Holstein--Primakoff transformation\cite{Holstein40} we obtain a bosonized spin Hamiltonian of the form\cite{KreiselEPJ}
\begin{equation}
  {\cal{H}} = {\cal{H}}_{0} + \sum_{n=2}^{\infty} {\cal{H}}_n \; .
 \end{equation}
Considering the large effective spin $S=M_\mathrm{s} a^3/\mu \approx 14.2$, it is sufficient to retain only terms up to $n=2$ in the formal $1/S$ expansion in order to calculate the magnetic excitation spectrum. The quadratic part of the Hamiltonian reads
 \begin{equation}
 {\cal{H}}_2 = \sum_{ij} \left[ A_{ij} b^{\dagger}_i b^{\phantom{\dagger}}_j + \frac{B_{ij}}{2}
 \left( b_i b_j + b^{\dagger}_i b^{\dagger}_j \right)  \right]\; ,
 \label{eq:H2}
\end{equation}
with
 \begin{subequations}
 \begin{align}
 A_{ij} & =  \delta_{ij} h + S ( \delta_{ij} \sum_{n} J_{ in} - J_{ ij} )
 \nonumber
 \\
 & \phantom{=}+  S \left[ \delta_{ij} \sum_{n} D_{in}^{zz}-
\frac{D_{ij}^{xx} + D_{ij}^{yy}}{2} \right],
 \\
 B_{ij} & =  - \frac{S}{2}  \left[ D_{ij}^{xx} - 2 i  D_{ij}^{xy} - D_{ij}^{yy} \right].
 \end{align}
 \end{subequations}
A thin film of thickness $d=Na$ is obviously not translational invariant in all spatial directions, such that a full Fourier transformation cannot be used. Instead we set $ {\bf R}_i = ( x_i , {\bf r}_i )$ with ${\bf r}_i = ( y_i , z_i )$ and use the property of the discrete translational invariance in the $y$ and $z$ directions to partially diagonalize ${\cal{H}}_2$ via a partial Fourier transformation. We expand the bosonic operator
\begin{equation}
 b_i = \frac{1}{\sqrt{N_{y} N_z}} \sum_{ {\bf k}} e^{ i  {\bf k} \cdot {\bf r}_i }
 b_{\bf k} ( x_i )\;,
 \end{equation}
where $N_y$ and $N_z$ is the number of lattice sites in the $y$ and $z$ direction. The Hamiltonian (\ref{eq:H2}) then reads
\begin{align}
 {\cal{H}}_{2} & =  \sum_{\bf k} \sum_{ x_i, x_j}
 \Bigl \{ [ \mathbf{A}_{ \bf k} ]_{ ij }  b^{\dagger}_{ \bf k}  (  x_i )  b^{\phantom{\dagger}}_{ \bf k}  (  x_j )
 \nonumber
 \\
&\hspace{-.5cm}    +
  \frac{ [ \mathbf{B}_{ \bf k} ]_{ ij } }{2}  b_{ \bf k}  (x_i )  b_{ -\bf k}  ( x_j )
+   \frac{ [ \mathbf{B}_{ \bf k} ]_{ ij }^{ \ast } }{2}  b^{\dagger}_{ \bf k}  ( x_i  )
b^{\dagger}_{ - \bf k}  ( x_j )
 \Bigr\},
 \label{eq:H2Gauss}
\end{align}
with the $N\times N$ matrices $\mathbf{A}_{ \bf k}$ and  $\mathbf{B}_{ \bf k}$ given by~\cite{Costa00}
\begin{subequations}
 \begin{align}
  [ \mathbf{A}_{ \bf k} ]_{ ij }   & =  \sum_{ \bf r}
 e^{ - i {\bf k} \cdot {\bf r}}  A ( x_i - x_j , {\bf r} )\; ,
 \nonumber\\
&=SJ_{\bf k}(x_{ij})+\delta_{ij}\bigl[ h+S\sum_{n}D_0^{zz}(x_{in})\bigr]\nonumber\\
&\phantom{=}-\frac S2 \bigl[D_{\bf k}^{xx}(x_{ij})+D_{\bf k}^{yy}(x_{ij})\bigr ] , \\
 [ \mathbf{B}_{ \bf k} ]_{ ij }  & = \sum_{ \bf r}
 e^{ - i {\bf k} \cdot {\bf r} } B ( x_i - x_j , {\bf r} )\nonumber\\
&\hspace{-.5cm}=-\frac S2 \bigl[D_{\bf k}^{xx}(x_{ij})-2i D_{\bf k}^{xy}(x_{ij})-D_{\bf k}^{yy}(x_{ij})\bigr].
 \end{align}
\label{eq:ab}
\end{subequations}
The exchange matrix is given by
\begin{align}
 J_{\bf k}(x_{ij})&=J \bigl[\delta_{ij}\bigl\{6-\delta_{j1}-\delta_{jN}\notag\\
&\hspace{-.5cm}-2(\cos (k_ya)+\cos(k_za))\bigr\}-\delta_{i j+1}-\delta_{i j-1}\bigr] ,
\label{eq:exmatrix}
\end{align}
and the dipolar tensor reads
\begin{align}
D_{\bf k}^{\alpha \beta} (x_{ij}) &=  {\sum_{{\bf r}_{ij}}}^\prime e^{-i {\bf k}\cdot {\bf r}_{ij}}  D_{ij}^{\alpha \beta} ,
\label{eq:dipsum}
\end{align}
where $\Sigma^\prime$ excludes the term  $y_{ij}=z_{ij}= 0$ when $x_{ij}=0$. We can rewrite the quadratic Hamiltonian in matrix notation
\begin{equation}
 {\cal{H}}_2=\frac 12 \sum_{\bf k} (\vec b_{\bf k}^\dagger,\vec b_{-\bf k}^{\phantom{\dagger}}) {\mathcal D}_{\bf k}\left(\begin{array}{c}
                                                                                              \vec b_{\bf k}\\
\vec b_{-\bf k}^\dagger
                                                                                             \end{array}\right)
 \label{eq:H2vec}
\end{equation}
with the grand dynamic matrix
\begin{equation}
 {\mathcal D}_{\bf k}=\left(\begin{array}{cc}
                             \mathbf{A}_{ \bf k}&\mathbf{B}_{ \bf k} \\
                             \mathbf{B}_{ \bf k}^T&\mathbf{A}_{ \bf k}^T \\
                            \end{array}\right)
\end{equation}
and the row vector
 \begin{equation}
  \vec  b_{\bf k}^\dagger =(b_{\bf k}^\dagger(x_1),\ldots,b_{\bf k}^\dagger(x_N))
 \end{equation}
Observing the low symmetry of the matrices
\begin{subequations}
\begin{align}
 \mathbf{A}_{ \bf k} = \mathbf{A}_{ \bf -k}^T\neq  \mathbf{A}_{ \bf k}^T\\
 \mathbf{B}_{ \bf k} = \mathbf{B}_{ \bf -k}^T\neq  \mathbf{B}_{ \bf k}^T
\end{align}
\end{subequations}
we have to use a full $2N$ square transformation
\begin{equation}
  {\mathcal J}_{\bf k}=\left(\begin{array}{cc} {\mathbf U}_{\bf k}^\dagger&- {\mathbf V}_{\bf k}^\dagger\\
- {\mathbf W}_{\bf k}^\dagger& {\mathbf X}_{\bf k}^\dagger\end{array}\right)
\end{equation}
to diagonalize the Hamiltonian (\ref{eq:H2vec}). This transformation connects the vectors of the true magnon operators $\vec{\gamma}_{\bf k}^\dagger=(\gamma_{{\bf k},1 }^\dagger,\ldots,\gamma_{{\bf k},N}^\dagger)$ to those in the Holstein--Primakoff basis via
\begin{equation}
 \left(\begin{array}{c}
        \vec{\gamma}_{\bf k}\\
\vec{\gamma}_{-\bf k}^\dagger
       \end{array}\right)={\mathcal J}_{\bf k}\left(\begin{array}{c}
        \vec{b}_{\bf k}\\
\vec{b}_{-\bf k}^\dagger
       \end{array}\right)=\left(\begin{array}{c}
      {\mathbf U}_{\bf k}^\dagger \vec{b}_{\bf k}^{\phantom{\dagger}}-{\mathbf V}_{\bf k}^\dagger \vec{b}_{-\bf k}^\dagger\\
-{\mathbf W}_{\bf k}^\dagger \vec{b}_{\bf k}+{\mathbf X}_{\bf k}^\dagger \vec{b}_{-\bf k}^\dagger
       \end{array}\right)\label{eq:trans}
\end{equation}
and transforms to the diagonal Hamiltonian
\begin{equation}
 {\cal{H}}_2=\frac 12 \sum_{\bf k}(\vec \gamma_{\bf k}^\dagger,\vec \gamma_{-\bf k}^{\phantom{\dagger}}) {\mathcal E}_{\bf k}\left(\begin{array}{c}
                                                                                              \vec \gamma_{\bf k}\\
\vec \gamma_{-\bf k}^\dagger
                                                                                             \end{array}\right)+E_0^{(2)}
\end{equation}
with the matrix
\begin{align}
  {\mathcal E}_{\bf k}&=\bigl(\mathcal J_{\bf k}^\dagger\bigr)^{-1} {\mathcal D}_{\bf k}^{\phantom{-1}}{\mathcal J}^{-1}_{\bf k}\notag\\
&=\mathop{\mathrm{diag}}(E_{{\bf k},1},\ldots,E_{{\bf k},N},E_{{\bf k},1},\ldots,E_{{\bf k},N})
\end{align}
and the correction
\begin{equation}
 E_0^{(2)}=\frac 12\sum_{\bf k}\mathop{\mathrm{Tr}} {\mathcal E}_{\bf k}
\end{equation}
to the groundstate energy.\\

In order to numerically calculate the eigenenergies $E_{{\bf k},n}$, $n=1,\ldots,N$ of all magnon modes and the transformation matrix for a given in-plane wavevector $\bf k$ we carry out the slowly converging sum in (\ref{eq:dipsum}) using Ewald summation technique to set up the grand canonical matrix ${\mathcal D}_{\bf k}$. Following the algorithm of Ref.~\onlinecite{Colpa78}, we first calculate the Cholesky decomposition $\mathcal K_{\bf k}$ of $D_{\bf k}$, which has the property
\begin{equation}
 \mathcal D_{\bf k}={\mathcal K}_{\bf k}^\dagger {\mathcal K}_{\bf k}^{\phantom{\dagger}}.
\end{equation}
Then we diagonalize the matrix ${\mathcal M}_{\bf k}$ given by
\begin{equation}
 {\mathcal M}_{\bf k}={\mathcal K}_{\bf k} I_p {\mathcal K}_{\bf k}\;,\quad I_p=\left(\begin{array}{cc}
                                                                                       1_N&0\\
                                                                                       0&-1_N
                                                                                      \end{array}\right)
\end{equation}
using a unitary transformation ${\mathcal U}_{\bf k}$ to obtain the diagonal matrix ${\mathcal L}_{\bf k}$ which is connected to the eigenenergies via
\begin{equation}
 {\mathcal E}_{\bf k}=I_p {\mathcal L}_{\bf k}.
\end{equation}
The inverse transformation
\begin{equation}
 {\mathcal J}^{-1}_{\bf k}=\left(\begin{array}{cc}
                                   {\mathbf U}_{\bf k}& {\mathbf W}_{\bf k}\\
                                   {\mathbf V}_{\bf k}&  {\mathbf X}_{\bf k}
                                 \end{array}\right)
\end{equation}
is now obtained from the equation
\begin{equation}
 {\mathcal J}^{-1}_{\bf k}={\mathcal K}_{\bf k}^{-1} {\mathcal U}_{\bf k}^{\phantom{-1}} {\mathcal E}_{\bf k}^{1/2}.
\end{equation}
Having constructed the full transformation, we then obtain the amplitude structure of the modes across the film. It is worth noting that in our approach the anti-nodal points of the magnon modes are located on the film surfaces which is consistent with the unpinned surface spin configuration.\cite{SW_and_pinning}

In line with the discussion in the previous section we also calculate the ellipticity of the modes $\epsilon_{{\bf k},n}$. The ellipticity can be calculated from the expectation values of the spin operators in the magnon eigenstates,\cite{Majlis}
\begin{equation}
 \epsilon_{{\bf k},n}=1-\frac{\langle (S^x)^2\rangle_{{\bf k},n}}{\langle (S^y)^2\rangle_{{\bf k},n}}.
\end{equation}
$S^\alpha$ denotes the $\alpha$ component of the total spin operator and the expectation value has to be taken in the magnon eigenstate $|{\bf k},n\rangle=\gamma_{{\bf k},n}^\dagger | 0\rangle$ of the $n$-th magnon mode. According to a calculation in lowest order $1/S$ we use the expansion of the Holstein--Primakoff transformation to express the spin operators in terms of boson operators to obtain
\begin{subequations}
\begin{align}
 \langle \bigl(S^{{x}}\bigr)^2\rangle_{{\bf k},n}&=\frac S4 \bigl[\langle \vec b_{-\bf k}\cdot \vec b_{\bf k}\rangle_{{\bf k},n}+ \langle \vec b_{-\bf k}^{\phantom{\dagger}}\cdot \vec b_{-\bf k}^\dagger\rangle_{{\bf k},n}\notag\\
&\phantom{=\frac S4 \bigl(}+\langle \vec b_{\bf k}^\dagger\cdot  \vec b_{-\bf k}^\dagger\rangle_{{\bf k},n}+  \langle \vec b_{\bf k }^\dagger\cdot  \vec b^{\phantom{\dagger}}_{\bf k }\rangle_{{\bf k},n}\bigr],\\
 \langle \bigl(S^{{y}}\bigr)^2\rangle_{{\bf k},n}&=\frac S4 \bigl[\langle \vec b_{-\bf k}\cdot \vec b_{\bf k}\rangle_{{\bf k},n}- \langle \vec b_{-\bf k}^{\phantom{\dagger}}\cdot \vec b_{-\bf k}^\dagger\rangle_{{\bf k},n}\notag\\
&\phantom{=\frac S4 \bigl(}+\langle \vec b_{\bf k}^\dagger\cdot  \vec b_{-\bf k}^\dagger\rangle_{{\bf k},n}-  \langle \vec b_{\bf k }^\dagger\cdot  \vec b^{\phantom{\dagger}}_{\bf k }\rangle_{{\bf k},n}\bigr].
\end{align}
\end{subequations}

An inversion of the transformation (\ref{eq:trans}) allows the calculation of the expectation values,
\begin{subequations}
 \begin{align}
  \langle \vec b_{-\bf k }^{\phantom{\dagger}} \cdot \vec b_{-\bf k}^\dagger\rangle_{{\bf k},n}&=\sum_j\bigl([U_{\bf k}^\dagger]_{nj}[U_{\bf k}^{\phantom{\dagger}}]_{jn}+[W_{\bf k}^\dagger]_{nj}[W_{\bf k}^{\phantom{\dagger}}]_{jn}\bigr),\\
  \langle \vec b_{\bf k }^\dagger \cdot \vec b_{\bf k }^{\phantom{\dagger}}\rangle_{{\bf k},n}&=\sum_j\bigl([V_{\bf k}^\dagger]_{nj}[V_{\bf k}^{\phantom{\dagger}}]_{jn}+[X_{\bf k}^\dagger]_{nj}[X_{\bf k}^{\phantom{\dagger}}]_{jn}\bigr),\\
\langle \vec b_{\bf k}^\dagger \cdot \vec b_{-\bf k}^\dagger\rangle_{{\bf k},n}&=\sum_j\bigl([U_{\bf k}^\dagger]_{nj}[V_{\bf k}^{\phantom{\dagger}}]_{jn}+[W_{\bf k}^\dagger]_{nj}[X_{\bf k}^{\phantom{\dagger}}]_{jn}\bigr),\\
\langle \vec b_{-\bf k} \cdot \vec b_{\bf k}\rangle_{{\bf k},n}&=\sum_j\bigl([V_{\bf k}^\dagger]_{nj}[U_{\bf k}^{\phantom{\dagger}}]_{jn}+[X_{\bf k}^\dagger]_{nj}[W_{\bf k}^{\phantom{\dagger}}]_{jn}\bigr),
 \end{align}
\end{subequations}
such that the ellipticity is given by
\begin{equation}
  \epsilon_{{\bf k},n}=\frac{2B_{{\bf k},n}}{A_{{\bf k},n}+B_{{\bf k},n}},
  \label{eq:quant_ellipticity}
\end{equation}
with the abbreviations
\begin{subequations}
\begin{align}
 A_{{\bf k},n}&=\sum_j\bigl(|[V_{{\bf k}}]_{jn}|^2+|[X_{{\bf k}}]_{jn}|^2 \notag\\
    &\phantom{= \sum_j\bigl(} +|[U_{{\bf k}}]_{jn}|^2+|[W_{{\bf k}}]_{jn}|^2\bigr),\\
 B_{{\bf k},n}&=2\mathop{\mathrm{Re}}\sum_j \bigl([V_{{\bf k}}]_{jn}[U_{{\bf k}}^\ast]_{jn}+[X_{{\bf k}}]_{jn}[W_{{\bf k}}^\ast]_{jn}\bigr).
\end{align}
\end{subequations}

\begin{figure}[tb]
\begin{center}
\scalebox{1}{\includegraphics[width=8.5 cm]{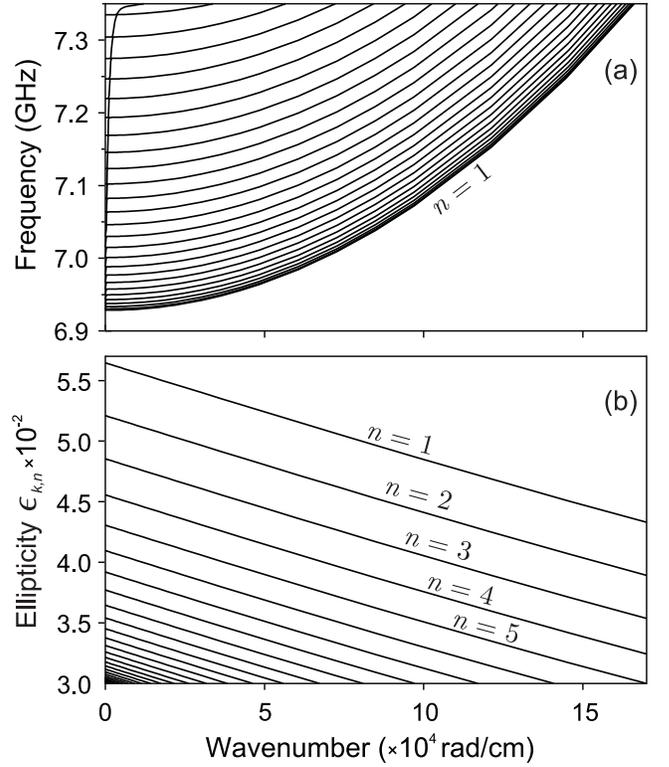}}
\end{center}
\caption{\label{figek} Theoretical result for the spectrum with $M_\mathrm{s}=1742\;\mathrm{G}$ (a) and the ellipticity of the corresponding magnon modes as a function of the wavevector (b) for a YIG film of thickness $d=5\;\mu\mathrm m$ in a magnetic field of $H_0=1750\;\mathrm{Oe}$. Our theoretical approach delivers the spectral position of all magnon modes for the model including both dipole-dipole and exchange interactions. The steepest section of the dispersion curves at small wavenumbers are the surface Damon-Eschbach mode which is strongly hybridized with the higher-order thickness modes (see e.g. Refs.~\onlinecite{Kozhus_1990, Kostylev_JMMM1995}).  Note that the ellipticity $\epsilon_{{\bf k},n}$ decreases as a function of the wavevector and with growing mode number $n$.}
\end{figure}

In the following we only analyze the characteristics of magnons with $\vartheta_{\bf{k}}={\pi}/{2}$ as these magnons have the highest ellipticity and consequently one may expect the strongest parametric coupling to this type of magnons. Furthermore, in our experiment these magnons propagate along the length of the microstrip transducer. This length is much greater than the free propagation path for the magnons. Therefore, one may expect that, in contrast to all other magnons which have a non-vanishing component of the in-plane wavevector perpendicular to the longitudinal axis of the transducer, the threshold of the parallel pumping excitation for them is not affected by the energy loss due to magnons escaping the pumped area.\cite{NeumannAPL09, Sholom} Figure~\ref{figek} shows the calculated magnon dispersion along with the ellipticity of the precession as a function of the magnon wavenumber $k$. The spectrum comprises the finite set (limited by the number of crystallographic elementary cells of $12.376$~{\AA} on the film thickness) of the quantized volume magnon modes (BVMSW) and a steep branch corresponding to the surface Damon-Eshbach magnon mode.

\section{Setup}
\label{sec:setup}

\begin{figure}[t]
\begin{center}
\scalebox{1}{\includegraphics[width=8.5 cm]{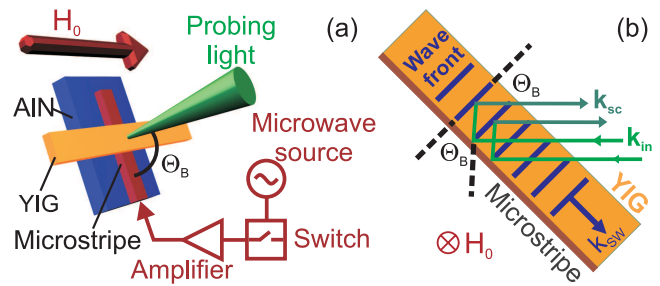}}
\end{center}
\caption{\label{Fig3} (Color online) (a) Geometry and block diagram for the experimental setup of the wave-vector selective BLS. (b) Geometry for the scattering of the probing light on the Bragg-grating generated by spin waves in the BLS technique.}
\end{figure}

The measurements were performed using combined microwave and optical facilities where the magnon system was parametrically pumped by the microwave circuit, and the response of the magnon system is analyzed by means of Brillouin light scattering spectroscopy. The microwave circuit comprises a microwave source, a switch, and a microwave amplifier connected to a probe section (see Fig.~\ref{Fig3}~(a)). In contrast to the conventional approach where the pumping frequency is held constant and the applied bias magnetic field $H_0$ is swept we examined the magnon spectrum at different pumping frequencies holding the applied field and consequently the ground state for small amplitude-excitations constant. This implies that the spectral characteristics shown in Fig.~2 were preserved during the experiment. The pumping microwave Oersted field was created by a 50~$\mu$m wide short-circuited microstrip line. The regular microstrip line was utilized instead of the commonly used microwave resonator~\cite{KurebayashiNatureMat11,Neumann08,NeumannAPL09} as it enables us to change the pumping frequency without the complicated realignment of the probe section. The investigated sample, a 15~mm long and 3~mm wide YIG film with a thickness of 5~$\mu$m, was placed on top of the microstrip line. An external static field $H_{0}$ of 1750~Oe was in the film plane and perpendicular to the longitudinal axis of the microstrip. In order to excite dipolar as well as exchange-dominated magnons, the frequency of the microwave source was varied from 13.6~GHz up to 14.6~GHz in steps of 20~MHz, which corresponds to change ${\omega_\mathrm{p}}/{2}$ from 6.8~GHz to 7.3~GHz in steps of 10~MHz. Thus, according to the magnon spectrum, a wide range of wavenumbers from zero up to $1.6\cdot 10^{5}$~rad/cm could be investigated by means of BLS spectroscopy.

The experiments were performed at different microwave powers from 100~mW to 10~W. In order to reduce any possible thermal effects which might have potentially influenced the magnon spectrum at high pumping powers by modifying the saturation magnetization $M_\mathrm{s}$, anisotropy fields, and the exchange stiffness constant the pumping was applied in 2~$\mu$s microwave pulses separated by 20~$\mu$s time intervals. For the same reason, a metallized aluminum nitride substrate, which is known for its high thermal conductivity, was chosen as a base plate for the probe section.

In order to probe the parametrically pumped magnons, the BLS measurements were performed in the back scattering geometry, where a single-mode solid state laser with a wavelength of 532~nm was focused onto the sample by using an objective with a high numerical aperture. The focal point of 20~$\mu$m in diameter was centered on the longitudinal axis of the microstrip line where the microwave Oersted field is parallel to the bias magnetic field $H_{0}$, and thus the condition for parallel parametric pumping is satisfied. The backscattered light was collected with the same objective and sent to a tandem Fabry-P\'{e}rot interferometer for frequency and intensity analysis. In a classical description, Brillouin light scattering can be interpreted as the diffraction of the probing light from a moving Bragg grating produced by a magnon mode, see Fig.~\ref{Fig3}~(b). As a result, some portion of the scattered light is shifted in frequency by the frequency of this mode (in our case ${\pm\omega_\mathrm{p}}/{2}$). The intensity of the inelastically scattered light is directly proportional to the number of magnons in the mode. In addition, the diffraction from the grating leads to a transfer of momentum during this process. By changing the angle $\Theta_\mathrm{B}$ between the sample and the incident light beam, the wavevector selection with a resolution of up to $4500$~rad/cm in a film plane can be implemented as shown in Ref.~\onlinecite{Sandweg}. The incident angle $\Theta_\mathrm{B}$ of the probing light determines the selected magnon wavenumber $k=2k_\mathrm{sc}\sin(\Theta_\mathrm{B})$, where the wavenumber of the scattered light $k_\mathrm{sc}$ is equal to the wavenumber of the incident light $k_\mathrm{in}$.

\section{Experimental results}
\label{sec:expdat}

\begin{figure}[tb]
\begin{center}
\scalebox{1}{\includegraphics[width=8.5 cm]{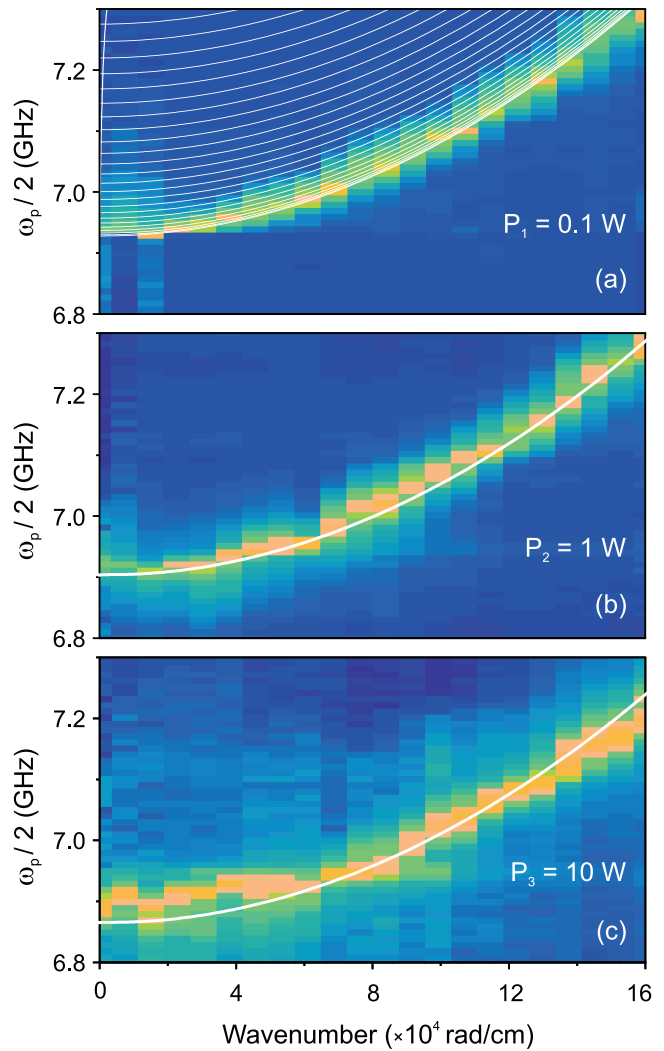}}
\end{center}
\caption{\label{Fig4} (Color online) Color-coded BLS-intensity map for the three different pumping powers of 10~W, 1~W and 100~mW. In panel (a) the white solid curves show the theoretically calculated dipolar-exchange spectrum of transversal spin waves propagating across the magnetization direction. In panels (b) and (c) the single white solid curves correspond to calculations of the lowest transversal magnon mode. The saturation magnetization has been used as a fit parameter: (a) $M_\mathrm{s}= 1742$~G, (b) $M_\mathrm{s}= 1715$~G, (c) $M_\mathrm{s}= 1676$~G. }
\end{figure}

Figure~\ref{Fig4} shows the measured BLS intensity of parametrically injected magnons as a function of the frequency and the wavevector for different pumping powers. To increase the signal-to-noise ratio the intensity was integrated across the entire width of the inelastically scattered peak for ${\omega_\mathrm{p}}/{2}$. The color-coded intensity maps, where blue (dark) corresponds to low intensities and orange (bright) to higher intensities, were recorded by changing the pumping frequency for the given incident angle $\Theta_\mathrm{B}$ of the probing light (see Fig.~\ref{Fig1}). Once the frequency measurement cycle had been completed, the incident angle $\Theta_\mathrm{B}$ was changed and the measurement were repeated in this way for wavenumbers ranging from 0 to $1.6 \times 10^{5}$~rad/cm in steps of $8.2 \times 10^{3}$~rad/cm.

In Fig.~\ref{Fig4}~(a), the color coded BLS-intensity map measured at the pumping power $P_{1}=0.1$~W, which is slightly (approximately 1~dB) above the threshold power of the parametric generation, is compared with the theoretical calculations of the relevant modes of the magnon spectrum. It is clearly visible that the detected magnons are located along the dispersion curve representing the lowest magnon mode $n=1$ while other regions of the spectrum show practically no increase in the BLS intensity. This behavior can be attributed to the fact that from all magnon groups pumped at the same time only the dominant group having the lowest damping and the highest coupling to the pumping field is significantly populated.

\begin{figure}[tb]
\begin{center}
\scalebox{1}{\includegraphics[width=8.5 cm]{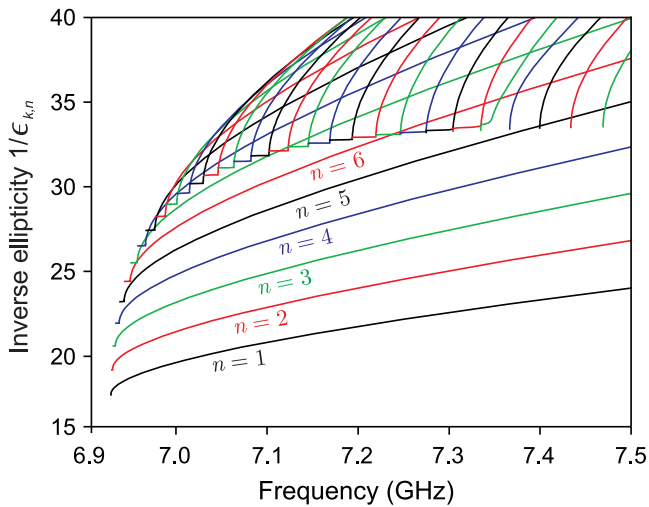}}
\end{center}
\caption{\label{Fig5} (Color online) Inverse ellipticity curves calculated as a function of frequency for the magnon modes propagating across the magnetization direction in a YIG film of thickness $d=5~\mu$m. The saturation magnetization $M_\mathrm{s}=1742$~G corresponds to the pumping power of $P_{1}=0.1$~W. Since the ellipticity $\epsilon_{{\bf k},n}$ is largest and the corresponding threshold of parametric generation $h_\mathrm{crit} \propto 1/\epsilon_{{\bf k},n}$ is lowest for the lowest magnon mode, $n=1$, pumping to this mode is expected at any given frequency.}
\end{figure}

The lowest mode corresponds to the backward volume magnetostatic wave having one node along the thickness of the YIG film. Modes with higher frequencies belong to higher order standing-wave modes, quantized perpendicular to the film plane. As we have no reason to assume different damping for different modes spread across the relatively narrow frequency and wavenumber ranges, our attention must be focused on the coupling efficiency. Since the elliptical precession of the magnetization allows one to excite the corresponding magnon modes (compare Fig.~\ref{Fig1}), as discussed above, we assume that the threshold of parametric generation is proportional to the inverse ellipticity, $h_\mathrm{crit} \propto 1/\epsilon_{{\bf k},n}$. Observing Fig.~\ref{Fig5}, one notices that the lowest magnon mode, $n=1$, has the largest ellipticity, and correspondingly the lowest threshold, in the entire pumping frequency range. Therefore, one would expect parametric amplification of this mode as the dominant magnon mode.

While the bias magnetic field is kept fixed during the measurement, the saturation magnetization $M_\mathrm{s}$ is influenced by the pumping of magnons. Further, single magnons carry a total spin of 1 and every magnon reduces the magnetization by one Bohr magneton $\mu_\mathrm{B}$. In order to account for that in the theoretical calculation of the spectrum we fix the exchange coupling $J$ at zero pumping and set the relevant spin stiffness to
\begin{equation}
 D=\frac{J M_\mathrm{s} a^5}{\mu^2}\;.
\end{equation}
Since the absolute magnon densities, composed by thermal magnons and pumped magnons, are unknown at this point, we use the saturation magnetization as a fit parameter. An adjustment of $M_\mathrm{s}$ then changes the position of the ferromagnetic resonance and the slope of the spectrum that originates from the effective exchange couplings. However, the slope of the spectrum also depends on the saturation magnetization and therefore needs to be adjusted to the experimental data. The solid curves in Fig.~\ref{Fig4}(b)-(c) correspond to the transversal BVMSW mode having the lowest energy. To fit the experimental data we varied the value of the saturation magnetization. We saw that with increasing pumping power, $M_\mathrm{s}$ decreased significantly, in accordance with the theory, from 1750~G for the undisturbed sample to 1676~G at pumping level of $P_{3}=10$~W. At the same time, one can see that the precision of the fit decreases for the largest pumping power. It can be connected to a spread of magnon population beyond the dominant magnon mode and to corresponding blurring of the BLS intensity over the wide frequency and $k$-number ranges.
This spread is caused by a number of physical phenomena. Firstly, at high pumping levels a number of frequency-degenerated magnon groups with damping higher than one for the dominant group can be excited.\cite{Lvov} Secondly, two-magnon scattering processes from crystal defects, dopants, and other static magnetic non-uniformities breaks the law of conservation of momentum, and thus lead to a spreading of the parametrically excited magnon population along the isofrequency lines in the ${\bf k}$-space. Thirdly, interactions between magnons and lattice vibrations will cause a redistribution of the magnons towards states with lower energy.~\cite{comment1} Finally, non-elastic four-magnon scattering processes are responsible for occupation of energy levels around the initial magnon state at half of the pumping frequency, and thus for consequent thermalization of the parametrically injected magnons across the entire spin-wave spectrum.~\cite{Cherepanov86}

In addition, we want also to comment on the role of the intermodal concurrency in parametric generation.\cite{Kozhus_experiment} We have observed no fine structure in the occupation of quantized standing modes at the beginning of the magnon spectrum as it was predicted by the theory in Refs.~\onlinecite{Kostylev_JMMM1995, Kozhus_theory} and reported in Ref.~\onlinecite{Wiese94}. This might be due to the fact that under our experimental conditions the inverse ellipticity curve of the lowest BVMSW mode lies below the corresponding curves of the highest standing modes and, as a result, this mode is solely amplified across the entire range of the experimentally probed frequencies (Fig.~\ref{Fig5}). Moreover, in accordance with our estimation based on Eq.~\ref{eq:quant_ellipticity}, the inverse ellipticity curve of the lowest BVMSW mode intersects first with the 65th standing mode at 9.6~GHz. The in-plane wavenumber of the lowest mode at this frequency is $k = 4\times 10^{5}$~rad/cm. The wavenumber of the 65th mode, calculated across the film, is $n \pi/d = 4.08\times 10^{5}$~rad/cm. Both of these values are significantly larger than the highest optically accessible wavenumber value of $2.6\times 10^{5}$~rad/cm. As a result no intermodal jumps in the magnon generation, and consequently no discrete structure in the BLS intensity map are expected to be observed in our experiment.

\section{Conclusion}
\label{sec:conclusion} We have observed the generation of parametric magnons under parallel pumping using the wavevector resolved BLS technique. In the theoretical part we constructed the magnon eigenstates for a thin ferromagnetic film in the framework of a microscopic Heisenberg model in linear spin-wave theory to be able to calculate such physical characteristics as eigenfrequencies, ellipticity, and spatial distribution of magnon modes. A good agreement between the experimentally determined spectral position of photon-coupled magnon pairs in a tangentially magnetized YIG film and the lowest-frequency magnon mode propagating in a film plane perpendicularly to the magnetization direction was obtained. Combining the theoretical results with the experimental observations, the dominant parametric excitation of the lowest mode is understood as a result of its highest ellipticity in the range of accessible wavenumbers. A fine structure which was previously reported in the threshold of parametric generation in YIG films was not detected in our experiment. This is possibly caused by the structure of the magnon spectra of the magnetic films in our case: the ellipticity of the lowest-frequency transversal magnon mode may be maximal across the entire range of the optically detectable $k$-numbers. In addition, a significant broadening of the populated spectral area is observed at high pumping powers. This phenomenon is associated with a variety of non-linear and linear scattering processes leading to thermalization of the parametrically injected magnons.

\begin{acknowledgments}
Financial support from the Deutsche Forschungsgemeinschaft within the SFB/TRR 49, from the Australian Research Council, and from the Australian-Indian Strategic Research Fund is gratefully acknowledged.
\end{acknowledgments}


\begin{thebibliography}{1}
\bibitem{comment1} It is worth to note that below the lowest transversal magnon mode, $n=1$, there are
    plenty of magnon states with a nonzero $k$-vector component along the magnetization direction.
\end{thebibliography}
\end{document}